\documentclass[runningheads]{llncs}
\usepackage[T1]{fontenc}
\usepackage{graphicx}
\usepackage{hyperref}       % hyperlinks
\usepackage{url}            % simple URL typesetting
\usepackage{booktabs}       % professional-quality tables
\usepackage{amsfonts}       % blackboard math symbols
\usepackage{nicefrac}       % compact symbols for 1/2, etc.
\usepackage{microtype}      % microtypography
\usepackage{graphicx}
\usepackage{amsmath}
\usepackage{booktabs}
\usepackage{multirow}
\usepackage{threeparttable}  % for table notes
\usepackage{subcaption}
\usepackage[font=small,labelfont=bf]{caption}
\usepackage{float}
\graphicspath{ {./images/} }
\usepackage{hyperref}

\makeatletter
\newcommand{\printfnsymbol}[1]{%
  \textsuperscript{\@fnsymbol{#1}}%
}
\makeatother
\usepackage{color}

\begin{document}
\title{Contrast-Invariant Self-supervised Segmentation for Quantitative Placental MRI}
\titlerunning{Contrast-Invariant Placenta Segmentation}

% \author{Anonymized Authors}  %% Added for anonymized MICCAI 2025 submission
% \authorrunning{Anonymized Author et al.}
% \institute{Anonymized Affiliations \\
%     \email{email@anonymized.com}
%     }
\author{
Xinliu Zhong \thanks{Equal contribution.} \inst{1,2} \and
Ruiying Liu \printfnsymbol{1} \inst{2} \and
Emily S. Nichols \inst{3,4} \and
Xuzhe Zhang \inst{5} \and
Andrew F. Laine \inst{5,6} \and
Emma G. Duerden \inst{3,4,7} \and
Yun Wang\thanks{Corresponding author.} \inst{1,2}
}

\authorrunning{Zhong et al.}

\institute{
Department of Computer Science, Emory University, Atlanta, GA 30307, USA \and
Department of Biomedical Informatics, Emory University, Atlanta, GA 30307, USA \and
Applied Psychology, Faculty of Education, Western University, London, ON N6A 3K7, Canada \and
Western Institute for Neuroscience, Western University, London, ON N6A 3K7, Canada \and
Department of Biomedical Engineering, Columbia University, New York, NY 10027, USA \and
Department of Radiology, Columbia University Irving Medical Center, New York, NY 10032, USA \and
Division of Maternal, Fetal and Newborn Health, Children’s Health Research Institute, London, ON N6C 2V5, Canada
}

\maketitle              % typeset the header of the contribution
\begin{abstract}
Accurate segmentation is critical for quantitative analysis of the placenta, yet remains challenging in T2*-weighted MRI due to echo-dependent contrast variation and limited manual annotations across echoes. We propose a contrast-augmented segmentation framework that exploits the inherent diversity of multi-echo T2*-weighted MRI to learn robust, contrast-invariant representations. Our method integrates: (i) masked autoencoding (MAE) for self-supervised pretraining on unlabeled multi-echo slices; (ii) masked pseudo-labeling (MPL) for semi-supervised domain adaptation across echo times; and (iii) global-local collaboration to align patch-level features with global anatomical context. We further introduce a semantic matching loss to encourage representation consistency across echoes of the same subject. Experiments on a clinical multi-echo placental MRI dataset demonstrate that our approach generalizes effectively across echo times and outperforms supervised baselines. To our knowledge, this is the first systematic framework tailored to multi-echo placental segmentation in T2*-weighted MRI.

\end{abstract}
\section{Introduction}
The placenta is vital for supporting fetal development by mediating oxygen, nutrients, and waste exchange between mother and fetus throughout pregnancy \cite{guttmacher2014human,regnault2002placental}. While ultrasound is the standard prenatal imaging tool, it can be limited by factors such as fetal position, maternal body habitus, or low amniotic fluid \cite{kline2010prenatal}. In contrast, magnetic resonance imaging (MRI) offers high soft-tissue contrast, a broad field of view, and multiplanar capabilities without ionizing radiation, making it a valuable alternative when ultrasound results are inconclusive. MRI is especially useful in evaluating complex placental conditions such as placenta accreta spectrum, where abnormal invasion of placental tissue into the uterine wall can lead to severe delivery risks.

Recent advances in quantitative MRI, such as relaxometry, allow in vivo assessment of placental function by capturing signal decay across multiple echo times and generating parametric maps (e.g., T1, T2, T2*) that reflect tissue composition and oxygenation \cite{haacke2015quantitative,zun2021feasibility}. While these biomarkers offer insights into placental development, their reliability hinges on accurate segmentation to define regions of interest.

Several deep learning methods have been proposed for placental segmentation, such as RFU-Net \cite{li2024placenta}, SegNeXt \cite{guo2022segnext}, and PLANET-S \cite{yamamoto2023planet}, all based on U-Net or CNN variants. However, these approaches primarily target anatomical MRI and offer limited exploration of quantitative imaging modalities. A recent study by Hall et al. \cite{hall2024placental} advanced the field by demonstrating automatic placental segmentation across multiple field strengths (0.55T–3T) and establishing normative T2* curves by gestational age. While this work confirms the feasibility of automated analysis in quantitative T2* MRI, it does not address segmentation across varying echo times within a single scan—where signal characteristics change rapidly due to contrast decay.  This setting presents unique challenges: placental appearance varies significantly across echo times, with boundaries that may blur or disappear due to contrast shifts. Moreover, annotations are typically available for only one or a few echoes, limiting supervised training. These factors reduce the effectiveness of standard models and highlight the need for contrast-invariant segmentation approaches.

In this work, we revisit multi-echo MRI as a form of contrast-augmented learning, hypothesizing that echo-wise contrast variation can serve as a natural self-supervised signal for learning robust, contrast-invariant representations. We propose a segmentation framework combining: (i) masked autoencoding for self-supervised pretraining on unlabeled multi-echo data, (ii) pseudo-labeling for unsupervised domain adaptation across echoes, and (iii) global-local consistency learning to align structural details with broader context. These components guide the model to focus on anatomy rather than contrast idiosyncrasies, improving generalization across echo times. We validate our method on a clinical multi-echo T2*-weighted placental MRI dataset and benchmark against multiple supervised baselines. Quantitative and qualitative results across eight echo times (TE1–TE8) demonstrate that our method consistently achieves superior segmentation performance, particularly under challenging later echoes. Notably, it maintains high Dice scores and low Hausdorff Distances even as image contrast degrades, underscoring its robustness to signal variability and echo-domain shifts.

\section{Methods}
\begin{figure}[h]
% \vspace{-5pt}
\centering
\begin{tabular}{cc}
    \begin{minipage}{0.496\textwidth}
        \centering
        \includegraphics[width=0.92\linewidth]{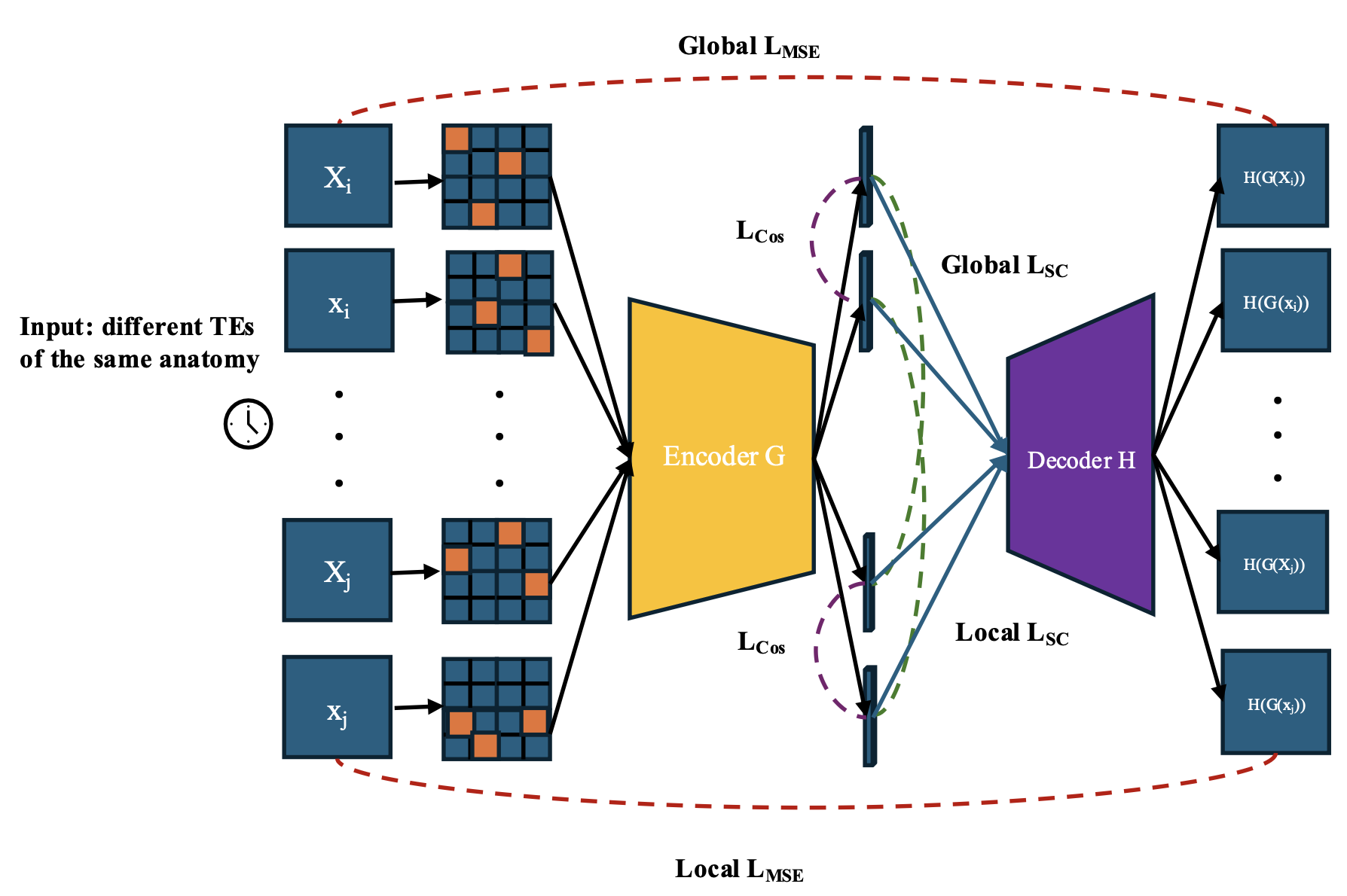}
        % \vspace{1mm}
        
        {\scriptsize \textbf{(a) MAE Pretraining.} Multi-echo slices are masked and encoded via a shared encoder. Training uses reconstruction loss ($\mathcal{L}_{\text{MSE}}$), semantic consistency ($\mathcal{L}_{\text{SC}}$), and global-local collaboration alignment ($\mathcal{L}_{\text{Cos}}$).}
    \end{minipage} &
    \begin{minipage}{0.49\textwidth}
        \centering
        \includegraphics[width=0.99\linewidth]{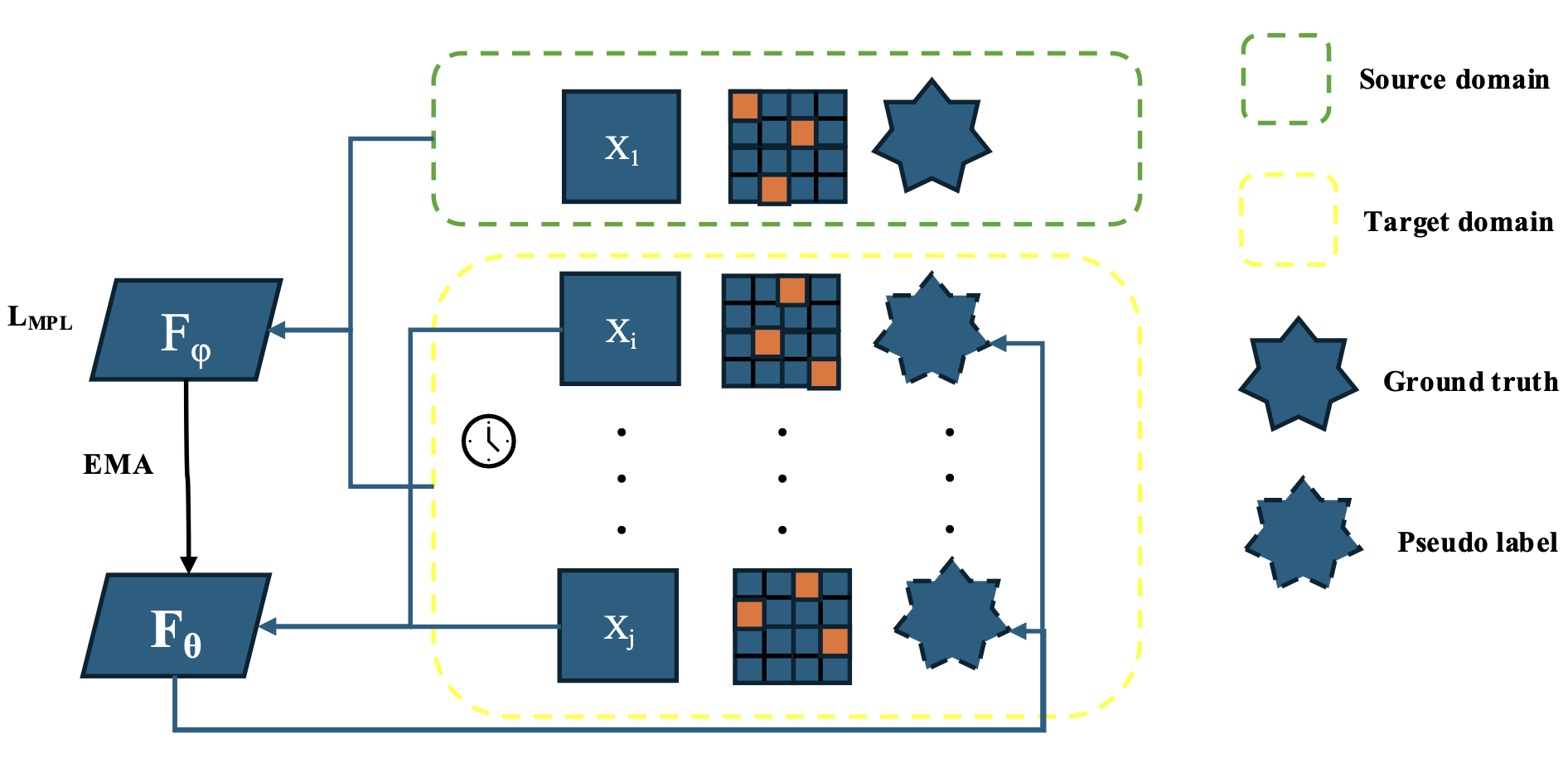}
        \vspace{0.03mm}
        
        {\scriptsize \textbf{(b) MPL Training.} A teacher-student setup leverages pseudo labels for unlabeled target slices. The teacher $F_\theta$ is updated by EMA, and the student $F_\phi$ is optimized with the MPL loss $\mathcal{L}_{\text{MPL}}$.}
    \end{minipage}
\end{tabular}
% \vspace{2mm}
\caption{\textbf{Overview of our two-stage training framework.} (a) Semantic MAE encourages robust contrast-invariant representations across echo times. (b) MPL enables segmentation adaptation using cross-domain pseudo supervision.}
\vspace{-10pt}
\label{fig:mae_mpl_framework}
\end{figure}
We propose a segmentation framework that maps multi-echo T2*-weighted placental images \( x_i \), each acquired at a distinct echo time \( \mathrm{TE}_i \in \{1, \ldots, N\} \), to corresponding segmentation masks \( y_i \). The framework tackles key challenges of fetal MRI—namely inter-echo contrast variation and limited supervision—via four integrated components:

\subsection{2D Masked Autoencoding}

To learn contrast-invariant anatomical representations without supervision, we adopt a 2D masked autoencoder (MAE \cite{he2022masked,zhang2024mapseg}). Each input slice is processed in two complementary forms: a local image patch \( x \) and its corresponding globally downsampled slice \( X \). Before being passed to the encoder, both are partially masked to produce \( x^M \) and \( X^M \), respectively.

The model is trained to reconstruct the original \( x \) and \( X \) from their masked versions, using a reconstruction loss over the missing regions:

\begin{equation}
    \mathcal{L}_{\text{MAE}} = \text{MSE}(g(x^M), x) + \text{MSE}(g(X^M), X),
\end{equation}

where \( g(\cdot) \) is the shared MAE encoder-decoder pipeline. This design encourages the encoder to capture semantic structure that is invariant to contrast differences across echo times, serving as a robust initialization for downstream tasks.
% This dual-scale training encourages the model to integrate fine-grained and global spatial cues, enabling robust pretraining across contrast-varying inputs.

\subsection{2D Pseudo-Labeling}
\label{sec:mpl}
To enable unsupervised domain adaptation across echo times, we extend the masked pseudo-labeling (MPL \cite{zhang2024mapseg}) framework to a 2D setting. After pretraining, the MAE encoder \( g \) is retained, and a segmentation decoder \( h \) is added to form the segmentation model \( f = h \circ g \).

We consider a labeled source domain slice \( x_s \) with its annotation \( y_s \), and an unlabeled target domain slice \( x_t \). Both undergo the same masking strategy used in pretraining, yielding \( x_s^M \) and \( x_t^M \). A teacher model \( f_\theta \) (with parameters \( \theta \)) predicts a pseudo-label \( \hat{y}_t = f_\theta(x_t) \) for the target image. The student model \( f_\phi \) is trained with a hybrid objective:

\begin{equation}
    \mathcal{L}_{\text{MPL}} = \mathcal{L}_{\text{Seg}}(f_\phi(x_t^M), \hat{y}_t) + \beta \mathcal{L}_{\text{Seg}}(f_\phi(x_s^M), y_s),
    \label{loss_mpl}
\end{equation}

where \( \beta \) controls the source-vs-target supervision ratio, and \( \mathcal{L}_{\text{Seg}} \) denotes a segmentation loss function (e.g., Dice or cross-entropy). The teacher parameters are updated using exponential moving average (EMA):

\begin{equation}
\theta_{t+1} \leftarrow \alpha \theta_t + (1 - \alpha)\phi_t.
\label{ema}
\end{equation}

This framework enables learning from unlabeled echoes by guiding the model with stable predictions, while still anchored to supervised TE1 annotations.

\subsection{Global-Local Collaboration}
\label{sec:glc}
To further improve segmentation robustness under domain shifts, we introduce a global-local collaboration \cite{zhang2024mapseg} module that fuses anatomical cues from different spatial scales. Given a local patch \( x \) and the downsampled slice \( X \), the encoder produces latent features \( \chi_{\text{loc}} = g(x) \) and \( \chi_{\text{glob}} = \text{upsample}(R \odot g(X)) \), where \( R \) denotes the location of \( x \) within \( X \) and \( \odot \) represents element-wise cropping. These are concatenated and passed to the decoder:
\begin{equation}
f(x) = h(\chi_{\text{loc}} \oplus \chi_{\text{glob}}),
\end{equation}
where \( \oplus \) denotes channel-wise concatenation. An auxiliary branch trained on \( X \) alone prevents overreliance on local features. We further regularize similarity between \( \chi_{\text{loc}} \) and \( \chi_{\text{glob}} \) via cosine similarity:

\begin{equation}
\mathcal{L}_{\text{cos}}(x, X) = 1 - \frac{\chi_{\text{loc}} \cdot \chi_{\text{glob}}}{\max(\|\chi_{\text{loc}}\|_2, \|\chi_{\text{glob}}\|_2, \epsilon)}.
\end{equation}

This loss is computed jointly on both source and target domains during the MPL stage and contributes to the overall training objective:
\[
\mathcal{L}_{\text{GLC}} = \gamma \, \mathcal{L}_{\text{seg}}(f(X), Y) + \delta \, \mathcal{L}_{\text{cos}}(x, X),
\]
where \( \mathcal{L}_{\text{seg}} \) is the supervised segmentation loss (e.g., Dice + CrossEntropy), and \( \gamma, \delta \) are hyperparameters. The total training loss is then:
\[
\mathcal{L}_{\text{total}} = \mathcal{L}_{\text{FSS}} + \mathcal{L}_{\text{MPL}} + \mathcal{L}_{\text{GLC}},
\]
where \( \mathcal{L}_{\text{FSS}} \) is the fully supervised source loss and \( \mathcal{L}_{\text{MPL}} \) is the masked pseudo-labeling loss defined in Section~\ref{sec:mpl}.

This encourages consistent feature representations across scales, improving resilience to spatial and contrast variability.

\subsection{Semantic Matching}

To ensure representation consistency across echo times (TEs), we introduce a semantic matching module that aligns latent features from the same anatomy under different contrast conditions. The core idea is that multi-echo images—despite visual differences—should yield semantically equivalent embeddings.

\paragraph{MAE Stage.} During pretraining, paired slices $(x_i, x_j)$ from the same spatial location but different TEs are passed through a shared encoder and decoder to produce features $z_i^{\text{enc}}, z_j^{\text{enc}}, z_i^{\text{dec}}, z_j^{\text{dec}}$. We define the semantic consistency loss as:
\begin{equation}
\mathcal{L}_{\text{SC}}^{\text{MAE}} = \lambda_{\text{enc}} (1 - \cos(z_i^{\text{enc}}, z_j^{\text{enc}})) + \lambda_{\text{dec}} (1 - \cos(z_i^{\text{dec}}, z_j^{\text{dec}})),
\end{equation}
% where $\cos(\cdot, \cdot)$ denotes cosine similarity.

\paragraph{MPL Stage.}  In the downstream MPL training, we apply the same feature alignment to the student model by minimizing:
\begin{equation}
\mathcal{L}_{\text{SC}}^{\text{MPL}} = \lambda_{\text{enc}}' (1 - \cos(z_i^{\text{enc}}, z_j^{\text{enc}})) + \lambda_{\text{dec}}' (1 - \cos(z_i^{\text{dec}}, z_j^{\text{dec}})).
\end{equation}

We omit KL-based distillation, as (1) hard pseudo-labels suffice, (2) identical teacher-student architectures reduce the need for soft-logit alignment, and (3) cosine loss offers a simpler, more stable alternative.

\section{Materials and Experiments}
\label{sec:materialsnexp}
\subsection{Dataset}
Placental multi-echo T2*-weighted MR images were collected from a prospective cohort at Western University (Dec. 2020–Aug. 2023). Scans were acquired using a 2D multi-echo gradient echo (FSPGR) sequence in oblique planes aligned with the placenta. Each acquisition captured 2–3 mid-placenta slices (slice thickness: 8.0 mm, spacing: 0.0 mm) with 8 echo times ranging from 3.15 to 37.45 ms (TR = 81.1 ms), using a flip angle of 30°. The in-plane resolution was approximately 1.37 × 2.73 mm² (matrix: 256 × 128, FOV: 35.0cm), and all echoes were reconstructed for downstream processing. No contrast agents were used.

% \paragraph{Patient Selection and Enrollment}

% A total of 54 pregnant individuals underwent one or two third-trimester MRI sessions (>=2 weeks apart): 53 in Session 1, and 44 returning for Session 2. Mean gestational ages were 31.8 weeks (S1) and 34.1 weeks (S2). The cohort included 22 female (40.7\%), 30 male (55.6\%), and 2 unknown-sex fetuses. Maternal age averaged 31.9 years (range: 22–41), with a mean socioeconomic score of 29.1 (SD = 12.0, scale: 0–45).
\paragraph{Cohort and Enrollment}
Fifty-four pregnant individuals underwent one or two third-trimester MRI sessions (>=2 weeks apart): 53 in Session 1 (mean GA: 31.8 wks), 44 in Session 2 (mean GA: 34.1 wks), with a GA range of 26.9 to 39.3 weeks. The cohort included 22 female (40.7\%), 30 male (55.6\%), and 2 unknown-sex fetuses. Maternal age averaged 31.9 years (range: 22–41), with a mean socioeconomic score of 29.1 (SD = 12.0, scale: 0–45).

\paragraph{Image Selection and Annotation}
For each scan, a senior radiologist selected two consecutive placenta-containing slices, yielding 211 annotated 2D multi-echo series. Manual segmentations were performed in FSLeyes (v1.6.1) and verified for anatomical accuracy.
% \paragraph{MRI Equipment and Parameters}

% Imaging was performed on a 3T GE Discovery MRI scanner using a 32-channel torso coil. Anatomical sequences were also acquired in axial, coronal, and sagittal planes using single-shot fast spin-echo protocols (TR > 1,000 ms, TE = 80 ms, 19–25 slices). In-plane resolution was 0.74 × 0.74 mm (coronal/sagittal) or 0.86 × 0.86 mm (axial), with a field of view (FOV) of 38–44 mm. Preprocessing included motion correction, bias field correction, and super-resolution reconstruction to 1 mm isotropic resolution. Each subject’s data was nonlinearly registered to a 36-week gestational age fetal brain atlas \cite{gholipour2017normative}, with atlas-based labels warped to native space.

\subsection{Data Splitting Strategy}
To promote contrast-invariant representation learning, we partitioned the dataset to separate MAE pretraining from MPL-based domain adaptation.

% We define \textbf{TE1} as the labeled source domain and treat all other echoes as part of a shared pool (\textit{TE\_other}) representing unlabeled target domains. The dataset is divided as follows:

\paragraph{MAE Pretraining Subset}
Five echo times (TEs) are randomly sampled from the unlabeled pool (\textit{TE\_other}), each contributing 211 slices. These $5 \times 211$ slices are used exclusively for MAE pretraining and excluded from downstream segmentation.

\paragraph{MPL Domain Adaptation}
TE1 serves as the labeled source domain, with 192 annotated slices. For the target domain, Two non-overlapping TEs from \textit{TE\_other} contribute $2 \times 211$ unlabeled slices. This setup simulates cross-contrast adaptation and allows evaluation of domain generalization. To assess robustness under varying contrast shifts, we define two specific target settings: (i) TE2 and (ii) TE6, each posing different levels of degradation.

\paragraph{Testing Protocol}
Five subjects are held out from the TE1 training pool, yielding 19 labeled slices for testing. These are fixed across all runs. All data splits are performed at the subject level to prevent any patient overlap between training, validation, or test sets, ensuring fair and leakage-free evaluation.

\subsection{Model Training and Testing}

Our training pipeline comprises two stages: (1) MAE for self-supervised representation learning, and (2) MPL for semi-supervised domain adaptation.

\paragraph{MAE Pretraining.}
Self-supervised pretraining was performed on unlabeled multi-echo slices. We used a 2D encoder-decoder with 8 layers and 512 embedding dimensions. Each input includes a local patch \(x\) and a downsampled global slice \(X\), both resized to $256 \times 256$. Patches of size \(8 \times 8\) (for \(x\)) and \(4 \times 4\) (for \(X\)) were randomly masked at 70\% to produce \(x^M\) and \(X^M\). The model was trained to minimize:
\begin{equation}
\mathcal{L}_{\text{MAE-total}} = \mathcal{L}_{\text{MSE}} + \gamma_{\text{SC}}^{\text{MAE}} \cdot \mathcal{L}_{\text{SC}}^{\text{MAE}}, \quad \gamma_{\text{SC}}^{\text{MAE}} = 0.4.
\end{equation}
Training used AdamW (lr = $2 \times 10^{-4}$, weight decay = 0.05, $\beta_1$ = 0.9, $\beta_2$ = 0.95), batch size 4, for 300 epochs.

\paragraph{MPL Domain Adaptation.}

The segmentation network \(f = h \circ g\) reuses the pretrained encoder \(g\) and adds a DeepLabV3-style decoder \(h\) \cite{chen2017rethinking}. Each batch includes a labeled TE1 slice \(x_s\) with label \(y_s\), and an unlabeled target slice \(x_t\); both are masked similarly to MAE. A pseudo-label \(\hat{y}_t = f_\theta(x_t)\) is produced by an EMA teacher \(f_\theta\), and the student \(f_\phi\) is trained with the total objective:
\begin{equation}
\mathcal{L}_{\text{MPL-total}} = \mathcal{L}_{\text{MPL}} + \mathcal{L}_{\text{GLC}} + \gamma_{\text{SC}}^{\text{MPL}} \cdot \mathcal{L}_{\text{SC}}^{\text{MPL}}, 
\quad \gamma_{\text{SC}}^{\text{MPL}} = 0.4.
\end{equation}
Here, \(\mathcal{L}_{\text{MPL}}\) is the hybrid Dice + cross-entropy loss as in Eq.~\ref{loss_mpl}, with \(\beta = 0.5\), and \(\mathcal{L}_{\text{GLC}}\) incorporates both global segmentation and cosine similarity regularization from Sec.~\ref{sec:glc}. The teacher parameters follow a staged EMA schedule: \(\alpha = 0.99\) (1k steps), 0.999 (2k), then 0.9999. Training ran for 150 epochs (including 50 warm-up), using AdamW (lr = $1 \times 10^{-4}$, weight decay = 0.01), batch size 1, and early stopping with 75-epoch patience.

\paragraph{Data Handling.}
Inputs are normalized to the 99.5th percentile with background exclusion. Augmentations (applied with 0.35 probability) include horizontal/vertical flips and intensity jittering.

% Both local and global branches were resized to $256 \times 256$ pixels during MAE and MPL training.

\paragraph{Training Infrastructure.}
All models were implemented in PyTorch and trained on NVIDIA H100 GPUs. 
% The MAE and MPL stages shared the same encoder to ensure consistent feature reuse.
% Checkpoints were saved every 50 epochs. Training progress and visualizations were tracked using Weights \& Biases (wandb). 

\section{Results and Discussion}

\subsection{Results}
\subsubsection{Quantitative Performance}

We evaluate segmentation performance using Dice Similarity Coefficient (Dice), Intersection over Union (IoU), pixel-wise Accuracy (Acc), Normalized Surface Dice (NSD), and Hausdorff Distance (HD). Dice, IoU, and Accuracy assess mask overlap, intersection quality, and classification correctness, respectively. To assess boundary precision, we include NSD and HD. NSD measures the proportion of surface points from both predicted and ground truth masks within a 1.0-voxel tolerance, providing a robust measure of contour agreement. HD quantifies the largest surface deviation, highlighting worst-case errors; lower values indicate fewer extreme misalignments.

Table~\ref{tab:all_te_results} presents the quantitative segmentation results across eight different echo times (TE1–TE8), comparing various baseline methods including U-Net \cite{ronneberger2015u}, U-Net++ \cite{zhou2018unet++}, LinkNet \cite{chaurasia2017linknet}, FPN \cite{lin2017feature}, DeepLabV3 \cite{chen2017rethinking}, nnU-Net \cite{isensee2021nnu} with our proposed approach. Our method consistently outperforms baselines, especially under challenging echo times (TE5–TE8), where contrast variation is more severe. It achieves the best or second-best Dice in 7 out of 8 TEs, with large gains at TE3 (92.5\%), TE5 (90.1\%), and TE6 (84.2\%) compared to the next best. Accuracy remains high consistently (>=92.5\%), peaking at 98.1\% for TE2. Moreover, HD is significantly reduced in most cases (e.g., 15.1mm at TE5 vs. 22–35mm for others), indicating better boundary alignment.

\begin{table*}[h]
\vspace{-13pt}
\centering
\caption{Segmentation performance across different echo times (TE1–TE8) for various methods. Metrics include Dice, IoU, Accuracy, and NSD (all in \%), and HD (in mm).}
\label{tab:all_te_results}
%\cite{ronneberger2015u} \cite{zhou2018unet++}  \cite{chaurasia2017linknet} \cite{lin2017feature} \cite{chen2017rethinking} \cite{isensee2021nnu}
\begin{subtable}{0.48\textwidth}
\centering
\scriptsize
\caption{TE1}
\resizebox{\textwidth}{!}{%
\begin{tabular}{lccccc}
\toprule
Method & Dice & IoU & Acc & NSD & HD \\
\midrule
U-Net  & 93.5\% & 88.5\% & 96.6\% & 89.0\% & 30.774 \\
U-Net++ & 95.4\% & 91.6\% & 97.7\% & 94.3\% & 12.412 \\
LinkNet  & 94.3\% & 89.8\% & 97.1\% & 90.4\% & 28.855 \\
 %Ma-net \cite{fan2020ma} & \textbf{95.2\%} & 91.2\% & 97.5\% & 93.8\% & 16.815 \\
FPN  & 95.1\% & 91.1\% & 97.6\% & 93.7\% & \textbf{11.524 }\\
DeepLabV3  & 94.3\% & 90.0\% & 97.0\% & 90.3\% & 21.143 \\
nnU-Net  & \textbf{95.8\%} & \textbf{92.3\%} & 97.8\% & \textbf{94.8\%} & 14.101 \\
Ours & 93.8\% & 88.4\% & \textbf{97.9\%} & 93.8\% & 12.726 \\
\bottomrule
\end{tabular}%
}\end{subtable}
\hfill
\begin{subtable}{0.48\textwidth}
\centering
\scriptsize
\caption{TE2}
\resizebox{\textwidth}{!}{%
\begin{tabular}{lccccc}
\toprule
Method & Dice & IoU & Acc & NSD & HD \\
\midrule
U-Net  & 93.6\% & 88.5\% & 96.7\% & 89.1\% & 32.244 \\
U-Net++  & 95.1\% & 91.2\% & 97.5\% & 93.8\% & 14.122 \\
LinkNet  & 94.6\% & 90.3\% & 97.3\% & 90.8\% & 29.568 \\
 %Ma-net \cite{fan2020ma} & \textbf{95.5\%} & \textbf{91.8\%} & 97.7\% & \textbf{94.4\%} & 12.589 \\
FPN  & 95.0\% & 91.0\% & 97.5\% & 93.6\% & 13.254 \\
DeepLabV3  & 94.5\% & 90.4\% & 97.2\% & 90.7\% & 22.002 \\
nnU-Net & \textbf{96.2\%} & \textbf{92.9\%} & \textbf{98.1\%} & 93.9\% & 13.441 \\
Ours & 94.2\% & 89.2\% & \textbf{98.1\%} & \textbf{94.2\%} & \textbf{12.435} \\
\bottomrule
\end{tabular}%
}\end{subtable}

\vspace{-6pt} 
\begin{subtable}{0.48\textwidth}
\centering
\scriptsize
\caption{TE3}
\resizebox{\textwidth}{!}{%
\begin{tabular}{lccccc}
\toprule
Method & Dice & IoU & Acc & NSD & HD \\
\midrule
U-Net  & 93.6\% & 88.5\% & 96.8\% & 89.1\% & 26.457 \\
U-Net++  & 90.0\% & 84.3\% & 95.7\% & 84.8\% & 22.530 \\
LinkNet  & 88.1\% & 82.8\% & 95.0\% & 79.2\% & 35.731 \\
 %Ma-net \cite{fan2020ma} & \textbf{93.7\%} & \textbf{89.2\%} & 97.0\% & 91.2\% & 16.259 \\
FPN  & 88.2\% & 82.3\% & 94.4\% & 82.0\% & 35.386 \\
DeepLabV3  & 90.9\% & 85.1\% & 95.5\% & 84.5\% & 27.302 \\
nnU-Net & \textbf{94.7\%} & \textbf{90.6\%} & 97.3\% & 92.1\% & 25.296 \\
Ours & 92.5\% & 86.4\% & \textbf{97.6\%} & \textbf{92.5\%} & \textbf{15.115} \\
\bottomrule
\end{tabular}%
}\end{subtable}
\hfill
\begin{subtable}{0.48\textwidth}
\centering
\scriptsize
\caption{TE4}
\resizebox{\textwidth}{!}{%
\begin{tabular}{lccccc}
\toprule
Method & Dice & IoU & Acc & NSD & HD \\
\midrule
U-Net  & 91.3\% & 85.3\% & 95.9\% & 85.0\% & 37.110 \\
U-Net++  & 89.6\% & 84.4\% & 96.1\% & 83.6\% & 24.057 \\
LinkNet  & 83.9\% & 78.2\% & 93.5\% & 71.6\% & 34.862 \\
 %Ma-net \cite{fan2020ma} & 87.8\% & 82.4\% & 95.3\% & 80.6\% & 25.738 \\
FPN  & 80.3\% & 74.0\% & 91.1\% & 67.7\% & 56.049 \\
DeepLabV3  & 85.3\% & 77.8\% & 93.8\% & 74.3\% & 37.151 \\
nnU-Net & 91.5\% & 84.2\% & 97.2\% & 87.9\% & 15.152 \\
Ours & \textbf{92.2\%} & \textbf{85.8\%} & \textbf{97.6\%} & \textbf{92.2\%} & \textbf{14.776} \\
\bottomrule
\end{tabular}%
}\end{subtable}

\vspace{-6pt} 
\begin{subtable}{0.48\textwidth}
\centering
\scriptsize
\caption{TE5}
\resizebox{\textwidth}{!}{%
\begin{tabular}{lccccc}
\toprule
Method & Dice & IoU & Acc & NSD & HD \\
\midrule
U-Net  & 86.3\% & 78.9\% & 94.2\% & 75.9\% & 29.527 \\
U-Net++  & 82.5\% & 75.9\% & 93.9\% & 70.8\% & 36.483 \\
LinkNet  & 78.8\% & 72.4\% & 91.7\% & 62.2\% & 42.214 \\
 %Ma-net \cite{fan2020ma} & 85.9\% & 80.2\% & 95.2\% & 76.7\% & 28.960 \\
FPN  & 74.7\% & 67.7\% & 88.8\% & 57.6\% & 67.182 \\
DeepLabV3  & 80.0\% & 71.6\% & 92.2\% & 64.4\% & 43.330 \\
nnU-Net & 90.0\% & 76.5\% & 95.8\% & 88.6\% & 26.944 \\
Ours & \textbf{90.1\%} & \textbf{82.4\%} & \textbf{97.1\%} & \textbf{90.1\%} & \textbf{17.250} \\
\bottomrule
\end{tabular}%
}\end{subtable}
\hfill
\begin{subtable}{0.48\textwidth}
\centering
\scriptsize
\caption{TE6}
\resizebox{\textwidth}{!}{%
\begin{tabular}{lccccc}
\toprule
Method & Dice & IoU & Acc & NSD & HD \\
\midrule
U-Net  & 79.8\% & 72.2\% & 92.5\% & 63.8\% & 36.842 \\
U-Net++  & 77.3\% & 69.8\% & 92.7\% & 61.2\% & 42.035 \\
LinkNet   & 74.6\% & 68.0\% & 90.7\% & 54.4\% & 49.722 \\
 %Ma-net \cite{fan2020ma} & \textbf{86.2\%} & \textbf{78.6\%} & 94.8\% & 77.8\% & 31.636 \\
FPN  & 66.3\% & 58.4\% & 85.8\% & 42.5\% & 91.449 \\
DeepLabV3  & 71.6\% & 62.9\% & 89.9\% & 48.8\% & 60.694 \\
nnU-Net & 80.1\% & 69.1\% & 92.4\% & 79.1\% & 44.263 \\
Ours & \textbf{84.2\%} & \textbf{73.5\%} & \textbf{95.7\%} & \textbf{84.2\%} & \textbf{25.849} \\
\bottomrule
\end{tabular}%
}\end{subtable}
\vspace{-6pt} 

\begin{subtable}{0.48\textwidth}
\centering
\scriptsize
\caption{TE7}
\resizebox{\textwidth}{!}{%
\begin{tabular}{lccccc}
\toprule
Method & Dice & IoU & Acc & NSD & HD \\
\midrule
U-Net  & 72.1\% & 64.6\% & 90.6\% & 49.3\% & 46.606 \\
U-Net++  & 74.7\% & 66.0\% & 90.7\% & 57.2\% & 45.966 \\
LinkNet  & 68.9\% & 62.3\% & 89.1\% & 43.9\% & 53.212 \\
 %Ma-net \cite{fan2020ma} & \textbf{79.6\%} & 71.0\% & 92.5\% & 65.8\% & 43.953 \\
FPN  & 59.3\% & 51.7\% & 83.5\% & 29.3\% & 100.400 \\
DeepLabV3  & 68.7\% & 59.6\% & 88.4\% & 44.0\% & 63.954 \\
nnU-Net & 72.5\% & 64.5\% & 90.4\% & 72.9\% & 53.843 \\
Ours & \textbf{78.7\%} & \textbf{66.8\%} & \textbf{94.4\%} & \textbf{78.7\%} & \textbf{38.741} \\
\bottomrule
\end{tabular}%
}\end{subtable}
\hfill
\begin{subtable}{0.48\textwidth}
\centering
\scriptsize
\caption{TE8}
\resizebox{\textwidth}{!}{%
\begin{tabular}{lccccc}
\toprule
Method & Dice & IoU & Acc & NSD & HD \\
\midrule
U-Net  & \textbf{76.5\%} & \textbf{67.7\%} & 91.8\% & 57.5\% & \textbf{40.644} \\
U-Net++  & 75.4\% & 65.3\% & 90.3\% & 59.0\% & 41.436 \\
LinkNet  & 67.1\% & 60.1\% & 88.6\% & 40.6\% & 63.213 \\
 %Ma-net \cite{fan2020ma} & 75.1\% & 66.4\% & 91.4\% & 57.4\% & 46.398 \\
FPN  & 54.1\% & 47.1\% & 81.7\% & 19.8\% & 107.631 \\
DeepLabV3  & 67.0\% & 57.2\% & 87.2\% & 41.3\% & 64.588 \\
nnU-Net & 75.4\% & 66.2\% & 86.1\% & 61.0\% & 72.149 \\
Ours & 70.9\% & 58.2\% & \textbf{92.8\%} & \textbf{70.9\%} & 42.820 \\
\bottomrule
\end{tabular}%
}\end{subtable}
\vspace{-15pt} 

\end{table*}

% Even without explicit domain pairing, our model still achieves a strong Dice score of 91.2\%, highlighting the robustness of the learned features. In contrast, MONAIfbs—trained directly on source-labeled data—underperforms significantly at 79.95\%, likely due to its limited capacity to generalize across varying echo contrasts.
\begin{figure}[h]
% \vspace{-22pt}
\centering
\setlength{\tabcolsep}{4pt}
\renewcommand{\arraystretch}{1}

\begin{tabular}{cccccc}
\includegraphics[width=0.145\textwidth]{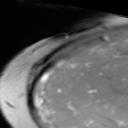} &
\includegraphics[width=0.145\textwidth]{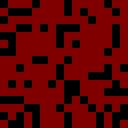} &
\includegraphics[width=0.145\textwidth]{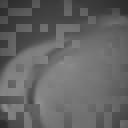} &
\includegraphics[width=0.145\textwidth]{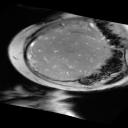} &
\includegraphics[width=0.145\textwidth]{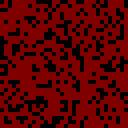} &
\includegraphics[width=0.145\textwidth]{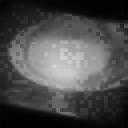} \\
 \((\mathrm{a})~x\) & 
 \((\mathrm{b})~M^{\text{x}}\) & 
 \((\mathrm{c})~\hat{x}^{\text{M}}\) & 
 \((\mathrm{d})~X\) & 
 \((\mathrm{e})~M^{\text{X}}\) & 
 \((\mathrm{f})~\hat{X}^{\text{M}}\)
\end{tabular}
\vspace{-6pt}  % tighter spacing before caption

\caption{
Qualitative results from MAE pretraining. (a–c): local patch input \(x\), mask \(M^x\), and reconstruction \(\hat{x}^M\). (d–f): corresponding results for the downsampled global slice. This figure demonstrates the model’s ability to recover multi-echo contrast features using masked self-supervised learning from both local and global views.
}
% \vspace{-2.3em}
% \vspace{-6pt}
\label{fig:mae_grid}
\end{figure}

\begin{figure}[h]
% \vspace{-10pt}
\centering
\setlength{\tabcolsep}{1pt}
\renewcommand{\arraystretch}{1.0}

% Top row (source domain)
\begin{minipage}[t]{0.14\textwidth}
  \centering
  \includegraphics[width=\linewidth]{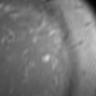}
  \vspace*{-19pt}
  \caption*{\((\mathrm{a})~x_{\text{s}}\)}
\end{minipage}
\hspace{4pt}
\begin{minipage}[t]{0.14\textwidth}
  \centering
  \includegraphics[width=\linewidth]{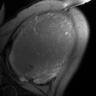}
  \vspace*{-19pt}
  \caption*{ \((\mathrm{b})~X_{\text{s}}\)}
\end{minipage}
\hspace{4pt}
\begin{minipage}[t]{0.14\textwidth}
  \centering
  \includegraphics[width=\linewidth]{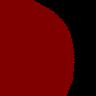}
  \vspace*{-19pt}
  \caption*{ \((\mathrm{c})~y_{\text{s}}\)}
\end{minipage}
\hspace{4pt}
\begin{minipage}[t]{0.14\textwidth}
  \centering
  \includegraphics[width=\linewidth]{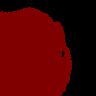}
  \vspace*{-19pt}
  \caption*{ \((\mathrm{d})~\hat{y}_{\text{s}}\)}
\end{minipage}
\hspace{4pt}
\begin{minipage}[t]{0.14\textwidth}
  \centering
  \includegraphics[width=\linewidth]{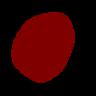}
  \vspace*{-19pt}
  \caption*{ \((\mathrm{e})~\hat{Y}_{\text{s}}\)}
\end{minipage}
\hspace{4pt}
\begin{minipage}[t]{0.14\textwidth}
  \centering
  \includegraphics[width=\linewidth]{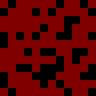}
  \vspace*{-19pt}
  \caption*{ \((\mathrm{f})~M_{\text{s}}\)}
\end{minipage}

\vspace{2pt} % vertical spacing between rows

% Bottom row (target domain)
\begin{minipage}[t]{0.14\textwidth}
  \centering
  \includegraphics[width=\linewidth]{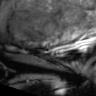}
  \vspace*{-19pt}
  \caption*{ \((\mathrm{g})~x_{\text{t}}\)}
\end{minipage}
\hspace{4pt}
\begin{minipage}[t]{0.14\textwidth}
  \centering
  \includegraphics[width=\linewidth]{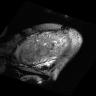}
  \vspace*{-19pt}
  \caption*{ \((\mathrm{h})~X_{\text{t}}\)}
\end{minipage}
\hspace{4pt}
\begin{minipage}[t]{0.14\textwidth}
  \centering
  \includegraphics[width=\linewidth]{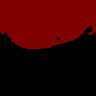}
  \vspace*{-19pt}
  \caption*{ \((\mathrm{i})~\tilde{y}_{\text{t}}\)}
\end{minipage}
\hspace{4pt}
\begin{minipage}[t]{0.14\textwidth}
  \centering
  \includegraphics[width=\linewidth]{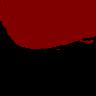}
  \vspace*{-19pt}
  \caption*{ \((\mathrm{j})~\hat{y}_{\text{t}}\)}
\end{minipage}
\hspace{4pt}
\begin{minipage}[t]{0.14\textwidth}
  \centering
  \includegraphics[width=\linewidth]{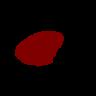}
  \vspace*{-19pt}
  \caption*{ \((\mathrm{k})~\hat{Y}_{\text{t}}\)}
\end{minipage}
\hspace{4pt}
\begin{minipage}[t]{0.14\textwidth}
  \centering
  \includegraphics[width=\linewidth]{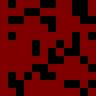}
  \vspace*{-19pt}
  \caption*{ \((\mathrm{l})~M_{\text{t}}\)}
\end{minipage}

\caption{
Qualitative results of MPL. Top row (a–f): source domain inputs including local/global views \(x_s, X_s\), ground truth \(y_s\), predictions \(\hat{y}_s, \hat{Y}_s\), and mask \(M_s\). Bottom row (g–l): target inputs \(x_t, X_t\), pseudo-label \(\tilde{y}_t\),  predictions \(\hat{y}_t, \hat{Y}_t\), and mask \(M_t\). This figure illustrates how the model leverages both views and semantic consistency to guide pseudo-label refinement in the absence of target-domain annotations. 
}
% \vspace{-10pt}
\label{fig:mpl_clean}
\end{figure}

\begin{figure}[htbp]
\centering
\includegraphics[width=\textwidth, height=0.42\textheight, keepaspectratio]{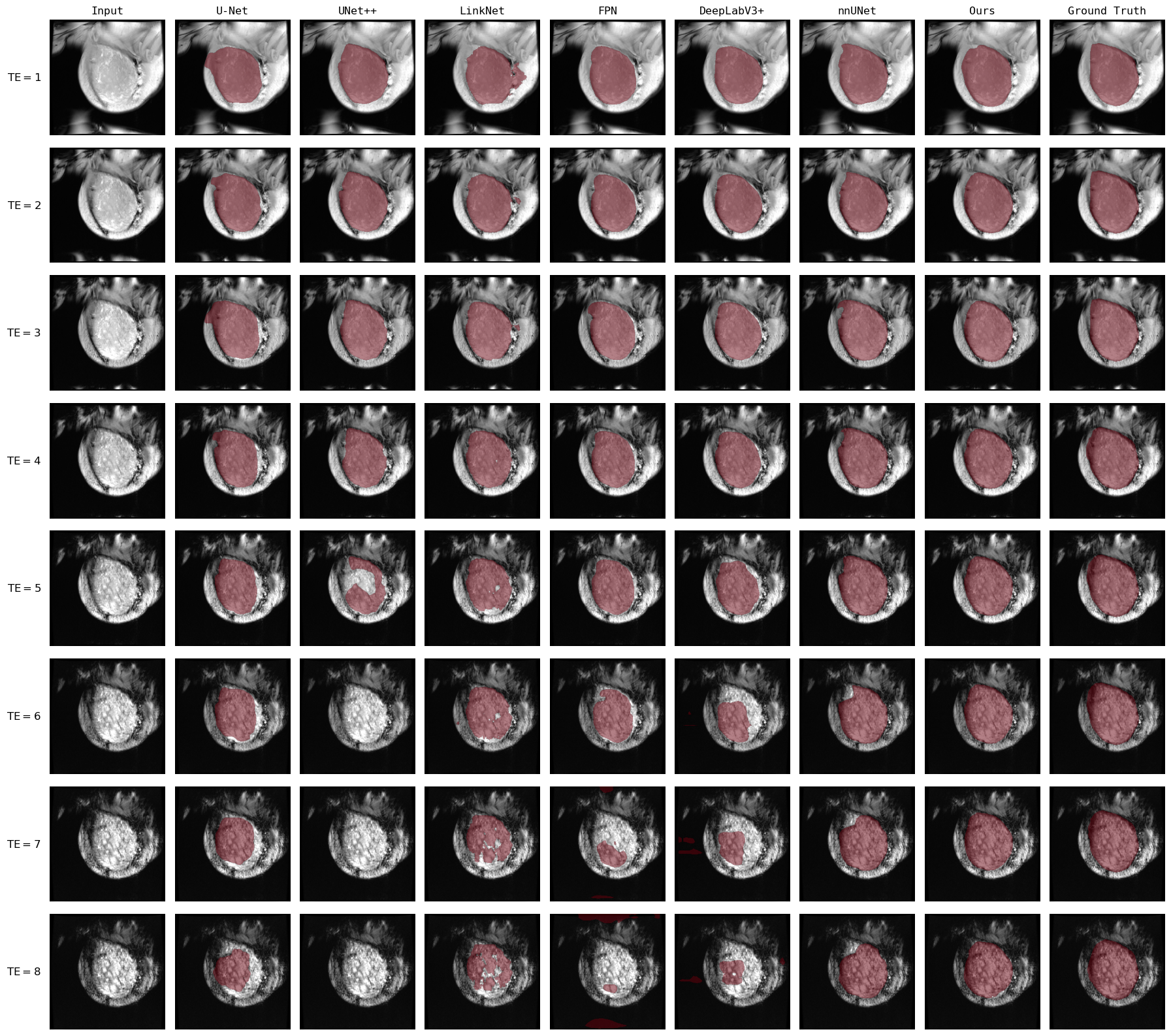}
\caption{Qualitative comparison of segmentation across echo times (TE1–TE8). Each row shows the input slice, model predictions (red overlay), and ground truth. Our method achieves consistent and accurate boundaries across varying contrast conditions.}
% \vspace{-50pt}
\label{fig:qualitative_te_comparison}
\end{figure}
\subsubsection{Qualitative Analysis}

Figure~\ref{fig:mae_grid} shows MAE reconstructions, demonstrating successful recovery of masked regions at both local and global scales, reflecting strong structural and contextual encoding. Figure~\ref{fig:mpl_clean} illustrates source and target predictions: the model aligns well with ground truth on source data and accurately segments target slices using pseudo-labels, indicating effective domain adaptation.

Figure~\ref{fig:qualitative_te_comparison} shows segmentation results across echo times (TE1–TE8). While all methods perform well at early echoes (TE1–TE3), baselines like U-Net and FPN degrade under low contrast at later echoes (TE6–TE8), with fragmented or blurry boundaries. In contrast, our method consistently yields smooth, anatomically aligned masks, especially at TE7 and TE8 where contrast variation is most severe. These visual trends support the quantitative results, confirming the model’s robustness to echo-domain shifts.

\subsection{Limitation}
Our current implementation is constrained to 2D networks due to the dataset structure, which contains only 2–3 annotated slices per subject. While this limits volumetric modeling, future extensions will explore 3D architectures once full-volume annotations become available. Manual labels were drawn on TE1 and reviewed for anatomical plausibility by trained raters, but further validation (e.g., by a second radiologist or on more challenging slices including fetal presence) is needed. We also note that our dataset only includes third-trimester cases (GA: 26.9–39.3 weeks), limiting generalizability to earlier gestational stages. Figures mainly show clear placental views, and the model’s robustness in more complex fetal-present fields of view remains to be evaluated.

% -When looking at the metrics, is there a statistically significant difference between the scores? It also appears that your method is not 1st or 2nd in two of the echos (TE1, TE2). discussion or limitation: baseline methods don't go through pretraining on other TEs, may overfit to TE1

% Can you explain the performance drop at TE8? UNet & UNet++ work as well or better here (which is against the intuition and idea of the paper). What do you think is causing this? --- claim differences between TE7 and TE8.

\subsection{Discussion}
Our two-stage training framework demonstrates strong segmentation performance across diverse echo times, confirming the value of contrast-invariant pretraining and domain-adaptive fine-tuning. Although the placenta remains largely stationary across echoes, we deliberately segment each echo independently to exploit multi-echo contrast diversity. Rather than copying a TE1-based label to other echoes, our method learns contrast-adaptive features that account for echo- and scanner-specific variations (e.g., T2* decay, B0 inhomogeneity, SNR differences). This strategy enriches the representation of tissue characteristics, supporting downstream applications such as pathology detection or quantitative biomarker extraction with minimal annotation overhead.

The strong quantitative and qualitative results confirm the effectiveness of our contrast-invariant segmentation framework. Notably, it achieves the highest Dice scores under challenging conditions (TE6–TE8), where baselines deteriorate due to low contrast and noise. For instance, while nnU-Net drops to 72.5\% Dice at TE7, our method retains a much higher 78.7\%, highlighting its robustness under domain shift. Substantial reductions in HD (e.g., 15.1mm at TE5, 17.5mm at TE7) indicate improved boundary delineation, enabled by global-local feature alignment and semantic consistency constraints. High NSD and Accuracy across echoes further confirm anatomical fidelity, even without target-domain labels.

It is worth noting that our method does not achieve top performance at TE1 and TE2, likely because competing baselines are directly trained on TE1 data, while our model prioritizes generalization across the echo spectrum. A minor performance drop at TE8 may reflect increased signal decay or artifact variability, suggesting a domain shift from earlier echoes. Finally, while the metrics show consistent improvement, statistical significance remains to be tested rigorously.

Overall, these findings validate the design of our two-stage training pipeline. MAE-based contrast-invariant pretraining enables the encoder to generalize across TEs, while MPL fine-tuning transfers supervision from TE1 to other echoes. The approach offers a scalable solution for echo-adaptive segmentation without 3D context or extensive annotation.

\section{Acknowledgement}
This work was supported by NIH grants R00HD103912 and R01MH133313 (Y.W.), and by the Canada First Research Excellence Fund, Canadian Institutes of Health Research, Molly Towell Perinatal Research Foundation, Brain Canada, and New Frontiers in Research Fund (E.N. and E.D.).
% \clearpage

\bibliographystyle{splncs04}
\bibliography{references}




% LNCS Homepage, \url{http://www.springer.com/lncs}, last accessed 2023/10/25
% \end{thebibliography}
\end{document}